\documentclass[12pt, a4paper]{article}
\usepackage{amsmath, amsthm,  amsfonts, amscd, epsfig}

\newcommand{\CO}{{\cal O}}

\newcommand{\CI}{{\cal I}}
\newcommand{\CL}{{\cal L}}
\newcommand{\CK}{{\cal K}}

\newcommand{\CS}{{\cal R}}

\newcommand{\db}{\overline{\partial}}
\newcommand{\Ga}{{\Gamma}}
\newcommand{\ind}{\mbox{ind}}
\newcommand{\be}[1]{\begin{equation} \label{#1}}

\newcommand{\p}{\partial}

\newcommand\tE{\tilde{E}}

\newcommand\longto{\longrightarrow }

\newcommand{\IP}{{\mathbb P}}
\newcommand\Z{\mathbb Z}
\newcommand\V{\bf V}
\newcommand\z{\zeta}
\newcommand\R{\mathbb R}
\newcommand\C{\mathbb C}

\renewcommand\O{{\mathcal O}}

\numberwithin{equation}{section}

\newcommand{\coker}{\text {coker}}

\theoremstyle{plain}
\newtheorem{theorem}{Theorem}[section]
\newtheorem{conjecture}[theorem]{Conjecture}
\newtheorem{lemma}[theorem]{Lemma}
\newtheorem{proposition}[theorem]{Proposition}

\theoremstyle{definition}
\newtheorem{definition}[theorem]{Definition}

\theoremstyle{remark}
\newtheorem{example}[theorem]{Example}

\title{On the complete integrability of the discrete Nahm equations}
\author{Michael K. Murray \\
Department of Pure Mathematics \\
University of Adelaide SA 5005\\
Australia
\and
Michael A. Singer\thanks{EPSRC Advanced Fellow}\\
Department of Mathematics and Statistics \\
James Clerk Maxwell
Building \\
University of Edinburgh EH9 3JZ \\
U.K.}

\begin{document}
\maketitle

\begin{abstract}
\noindent The discrete Nahm equations, a system of
matrix valued difference equations, arose in the work of
Braam and Austin on half-integral mass hyperbolic monopoles.

We show that the discrete Nahm equations are completely integrable in
a natural sense: to any solution we can associate a spectral curve and
a holomorphic line-bundle over the spectral curve, such
that the discrete-time DN evolution corresponds to walking in the
Jacobian of the spectral curve in a straight line through the line-bundle with steps of a fixed size.  Some of the implications for
hyperbolic monopoles are also discussed.

\vspace{12pt}

\noindent AMS Classification scheme numbers 39A12, 58F07
\end{abstract}

\section{Introduction}

This paper is concerned with two closely related stories: one about
the complete integrability of a discrete-time system of nonlinear
matrix equations (the discrete Nahm or DN system), the other having to
do with $SU_2$-monopoles on hyperbolic three-space $H^3$.  The link
between these two stories is given by the Braam--Austin version of the
ADHMN construction, which is a correspondence between hyperbolic
monopoles (of integral or half-integral mass)  and certain solutions
of the DN system \cite{BA}.

We shall show, using methods very close to those of \cite{NJH1}, that
the DN system is completely integrable in a natural sense: to any
solution we can associate a spectral curve $S$ and a holomorphic
line-bundle $\CL \to S$, such that the discrete-time DN evolution
corresponds to walking in the Jacobian of $S$ in a straight line
through $\CL$ with steps of a fixed size. The main novelty in this is
that $S$ lies in $\IP_1\times \IP_1$ rather than in the total space of
$\CO(d) \to \IP_1$.  It turns out that the geometry of $\IP_1\times
\IP_1$ gives rise in an entirely natural way to a discrete-time
system.  At the technical level, the new geometric set-up means that
it is necessary to develop a number of modifications of the modern
theory of algebraically integrable systems (by which we mean the body
of knowledge that is surveyed, for example, in \cite{NJH2}).

This account is accessible to readers with no knowledge of (or
interest in) hyperbolic monopoles. On the other hand the origin of the
DN system in the theory of hyperbolic monopoles provided us with
essential insights in this work, and is probably the main reason for
its interest.  Therefore we have also described how the particular
solutions of the DN systems that are linked by Braam and Austin to
hyperbolic monopoles arise within our general framework.  This leads
in particular to constraints on spectral curves of hyperbolic
monopoles analogous to those previously known in the euclidean case.

\vspace{10pt}
\noindent {\bf Acknowledgement} It is a pleasure to thank
Nigel Hitchin, Jacques Hurtubise, Antony Maciocia and Richard Ward for
a number of
useful conversations. An EPSRC Visiting Research Fellowship, which
allowed the second author to visit the first author for three months
in 1998 is also gratefully acknowledged.

\subsection{Nahm equations} The Nahm equations comprise the following
non-linear system of ordinary differential equations:
\be{eq1}
\frac{dT_1}{dz} = [T_2,T_3],\; \frac{dT_2}{dz} = [T_3,T_1],\;
\frac{dT_3}{dz} = [T_1,T_2]
\end{equation}
where the $T_i$ are functions of the real variable $z$, with values in
the complex, skew-hermitian $k\times k$ matrices. They form a
completely integrable
system which reduces, when $k=2$, to the Euler top equations. In
particular there is a Lax formulation
\be{eq2}
\frac{dA}{dz} = [A ,A_+]
\end{equation}
where
\be{eq3}
A:= A(\z) = (T_1+iT_2) - 2iT_3\z + (T_1 - iT_2)\z^2,\;\;
A_+:= A_+(\z) = -iT_3 + (T_1 - iT_2)\z.
\end{equation}
The complete integrability is obtained from this by setting up the
eigenvalue problem
\be{eq4}
A(\zeta)f(\eta,\zeta) = \eta f(\eta,\zeta).
\end{equation}
Then if $A(\zeta)$ evolves according to (\ref{eq2}) and
\be{eq5}
\frac{df}{dz} + A_+f =0,
\end{equation}
the eigenvalue $\eta$ remains constant. In particular the equation
\be{eq6}
\det(\eta - A(\z)) = 0
\end{equation}
must be independent of $z$, so the coefficients of this equation are a
set (in fact a complete set) of conserved quantities for the system
(\ref{eq1}).

For some purposes (and in particular to allow an easy comparison with
the discrete Nahm system to be introduced below) it is useful to
reformulate these equations slightly. First one introduces a `gauged'
version by adding a further skew-hermitian matrix-valued function
$T_0$, and writing
\be{eq7}
\frac{dT_1}{dz}-[T_0,T_1] = [T_2,T_3],\; \frac{dT_2}{dz}-[T_0,T_2] =
[T_3,T_1],\; \frac{dT_3}{dz}-[T_0,T_3] = [T_1,T_2].
\end{equation}
If one regards $d/dz -T_0$ as a connection, then (\ref{eq7}), modulo
gauge equivalence, is equivalent to the original system  (\ref{eq1})
(modulo conjugation by constant matrices).  To be quite explicit, the
gauge group here is
the space of smooth maps $g(z)$ into $U(k)$ and
\be{eq8}
g(T_0,T_i) = (gT_0g^{-1} - (dg/dz)g^{-1}, gT_ig^{-1}).
\end{equation}
Now, following Donaldson \cite{D}, we introduce the `complex
variables' $\sigma
= T_0 + iT_1$, $\tau = T_2 + iT_3$.  Then  (\ref{eq7}) becomes
\be{eq9}
\frac{d\tau}{dz} = [\sigma,\tau],\;
\frac{d\tau^*}{dz} = - [\sigma^*,\tau^*];
\end{equation}
and
\be{eq10}
\frac{d}{dz}(\sigma +\sigma^*) = [\sigma,\sigma^*] + [\tau,\tau^*].
\end{equation}

\subsection{Discrete Nahm equations}
\label{ba}

In \cite{BA} Braam and Austin found a discrete version of (\ref{eq1}).  The
relation of this system to the theory of monopoles on hyperbolic space
will be described in \S\ref{inst}. For the moment, let us just write
it down: \be{eq11} \beta_i\gamma_{i+1} = \gamma_{i+1}\beta_{i+2}, \;\;
\beta^*_{i+2}\gamma^*_{i+1} = \gamma^*_{i+1}\beta^*_{i}
\end{equation}
and
\be{eq12}
\gamma^*_{i-1}\gamma_{i-1} - \gamma_{i+1}\gamma^*_{i+1} +
[\beta_i^*,\beta_i]= 0.
\end{equation}
Here the discrete variable $i$ runs over $I= \{a,a+2,\ldots,b\}\subset
 2\Z$ and the $\beta$'s and
$\gamma$'s are $k\times k$ complex matrices, with the $\gamma$'s
invertible.  There is a gauge group $G$ which consists of sequences
$(g_i)$ of unitary matrices,  acting as follows:
\be{eq13}
\beta_i \mapsto g_i\beta_i g_i^{-1},\;\;
\gamma_{i+1} \mapsto g_i\gamma_{i+1}g_{i+2}^{-1}.
\end{equation}
This system of equations was also supplemented by a boundary condition
which we shall consider later.

The equations have a formal similarity to the standard Nahm system,
with (\ref{eq9}) and (\ref{eq10}) resembling, respectively
(\ref{eq11}) and (\ref{eq12}).  One aspect of this is that the latter
really are a discretization of the former. To see this, rescale $I$ by
multiplying by $h$ (which is to be thought of as small and
positive). Given $\sigma$ and $\tau$, set
\be{eq14}
\gamma_{i+1}^* = \frac{1}{2h} + \sigma(h(i+1)),\;
\beta_{i}^* = \tau(hi).
\end{equation}
Then we have
\be{eq15}
\beta_{i+2}^*\gamma^*_{i+1} - \gamma_{i+1}^*\beta^*_i
= \left[\frac{d\tau}{dz} - [\sigma,\tau]\right]_{z = ih} + O(h)
\end{equation}
and
\be{eq16}
\gamma^*_{i-1}\gamma_{i-1} - \gamma_{i+1}\gamma^*_{i+1} +
[\beta_i^*,\beta_i]
= -\left[\frac{d}{dz}(\sigma +\sigma^*) - [\sigma,\sigma^*] +
[\tau,\tau^*]\right]_{z =ih} + O(h),
\end{equation}
so that (\ref{eq11}) and (\ref{eq12}) are satisfied to lowest order in
$h$ by virtue of (\ref{eq9}) and (\ref{eq10}).

Note further that it is reasonable to think of the Braam--Austin equations
as the evolution equations of a discrete-time system.  For given
$\gamma_{i-1}$ and $\beta_i$,  we solve (\ref{eq12}) for
$\gamma_{i+1}$ and then (\ref{eq11}) determines $\beta_{i+2}$.  This
procedure gives a unique evolution (up to gauge) provided that the
quantity $\gamma^*_{i-1}\gamma_{i-1} + [\beta_i^*,\beta_i]$ is
positive-definite.  If this fails, then the evolution cannot  be
continued beyond this point.

We remark also that there is a natural way to fix the gauge
by taking $\gamma_{i+1}>0$ to be the (positive)
square root of $\gamma^*_{i-1}\gamma_{i-1} + [\beta_i^*,\beta_i]$, at
every step of the evolution: in other words, we take the $\gamma_i$ to
be self-adjoint. Comparing with (\ref{eq14}), we see that this
corresponds to the gauge $T_0=0$ and so to the original form
(\ref{eq1}) of the Nahm equations.

\subsection{Statement of results}
\label{sr}
The main purpose of this paper is to explain that the Braam--Austin
system also shares a more profound property with the standard Nahm
equations, their complete integrability.  In order to state our
results more precisely we must give a minor reformulation the Braam--Austin
equations.

First of all let us complexify the system, replacing $\beta$ by $-A$,
$\beta^*$ by $D$, $\gamma^*$ by $P^+$ and $\gamma$ by $-P^-$. (The
choice of signs is for later convenience only.)  We replace the index
set $I$ by a set $Z= \{r_0,r_0+1,\ldots,r_1-1,r_1\}$ of consecutive
integers ($r_0 \geq -\infty$, $r_1 \leq +\infty$).  We assume given a
complex $k$-dimensional vector space $V_r$ attached to each $r\in Z$
and naturally interpret $A_r$ and $D_r$ as endomorphisms of $V_r$.  By
contrast $P^+$ and $P^-$ map adjacent vector spaces to each other and
we shall choose the numbering so that $P^+_r$ maps $V_r$ to $V_{r+1}$,
while $P^-_r$ maps $V_{r}$ to $V_{r-1}$.

Now by {\em discrete Nahm data} at $r\in Z$, we mean a triple
$(A_r,B_r,D_r)$ of endomorphisms of $V_r$.  Given discrete Nahm data
at adjacent points $r$ and $r+1$ in $Z$ and maps $P^+_{r}:V_r \to
V_{r+1}$, $P_{r+1}^-:V_{r+1}\to V_r$, we say that the discrete Nahm
(DN) equations  are satisfied on $[r,r+1]$ if the following hold:
\be{eq2155}
P_{r+1}^-A_{r+1} - A_rP_{r+1}^- =0,\;\;
P_{r}^+D_{r} - D_{r+1}P_{r}^+ =0
\end{equation}
and
\be{eq2156}
B_{r} = P^-_{r+1}P^+_{r} + A_{r}D_r,\;\;
B_{r+1} = P^+_rP^-_{r+1} + D_{r+1}A_{r+1}.
\end{equation}
Furthermore we shall say that the DN equations are
satisfied on $Z$ if for {\em every} pair of adjacent points $r,r+1$ in
$Z$, the DN equations are satisfied on $[r,r+1]$.  It is clear that
(\ref{eq2155}) corresponds to (\ref{eq11}) and that if the second of
(\ref{eq2156}) holds with $r$ replaced by $r-1$, then we have at $r$
$$
B_{r} = P^-_{r+1}P^+_{r} + A_{r}D_r = P^+_{r-1}P^-_{r} + D_{r}A_{r}
$$
which yields (\ref{eq12}).  Thus (\ref{eq2155}) and (\ref{eq2156})
provide a reformulation of the Braam--Austin system except at the
end-points of $Z$. As in (\ref{eq13}) there is a natural gauge freedom
given by the action of $g_r \in {\rm GL}(V_r)$ where $g_r$ acts by
conjugation on the triple $(A_r,B_r,D_r)$ and by $P^{\pm}_r \mapsto
g_{r\pm 1}P^\pm_r g^{-1}_r$.

As well as taking care of the end-points, the introduction of $B$
allows us to define
the {\em spectral curve} $S$ of DN data. Given $(A_r,B_r,D_r)$
consider
\be{eq2158}
S_r = \{\det(\eta\zeta\, A_r  + \eta\, B_r + \zeta + D_r) = 0\}.
\end{equation}
This defines an algebraic curve in $\C^2$ which has a natural
compactification in $\IP_1\times\IP_1$ and which is gauge-independent.
The data also define a
holomorphic line-bundle\footnote{Strictly we should assume that $S_r$ is
smooth here} $\CL_r$ over $S_r$ as the cokernel of the
multiplication map
 $$
 \C^k\otimes\CO(-1,-1)  \xrightarrow{M_r(\eta,\zeta)}
 \C^k\otimes \CO
 $$
where $M_r$ is the matrix in (\ref{eq2158}).   Our main results are as
follows:
\begin{theorem} Given a solution $(A,B,D,P^{\pm})$ of the DN system
(\ref{eq2155}) and (\ref{eq2156}) in $Z$, we have $S_r = S_{r'}$ and
$\CL_{r'} = \CL_r\otimes L^{r'-r}$ for all $r,r' \in Z$.
\label{th1}\end{theorem}

Thus the spectral curve is constant for the DN evolution and that
evolution corresponds to walking in a straight line on the Jacobian of
$S$, with steps of fixed size, corresponding to the line-bundle $L=
\CO(1,-1)$ over $\IP_1\times \IP_1$.

The converse of Theorem~\ref{th1} is as follows:
\begin{theorem} \label{th2} Let $S$ be a smooth curve of bidegree
$(k,k)$ in
$\IP_1\times\IP_1$ and let $\CL$ be a regular holomorphic line-bundle
over $S$. Then there is canonically associated to $(S,\CL)$ a solution
$(A,B,D,P^{\pm})$ of the DN equations over $Z = \{r_0,\ldots r_1\}$,
such that the spectral curve of the solution is $S$ and $\CL_r \otimes
\CL^{-1}$ is an integral power of $L$, for every $r$ in $Z$.  The set
$Z$ is determined by the condition: $r\in Z$ if and only if for all
integers $m$  between $0$ and $r$ (inclusive), $\CL\otimes L^m$ is
regular.
\end{theorem}

The term `regular' is defined below (Definition~\ref{reg}); the set of
regular elements is a dense open subset of the Jacobian of $S$.

Combining these two Theorems we obtain a further result about the
evolution from initial data of the DN system.

\begin{theorem}
\label{th3}
 Let $(a,b,d)$ be a triple of $k\times k$ matrices such
that
$$
\{\det(\eta\zeta\, a + \eta\, b  + \zeta + d)=0\}
$$
is smooth in $\IP_1\times\IP_1$ and let $\CL \to S$ be defined as
above.  Then there exists a unique solution
$(A,B,D,P^{\pm})$ of the DN system on $Z$ such that $Z$ contains $0$
and $(A_0,B_0,D_0)$ is gauge-equivalent to $(a,b,d)$.  Moreover
$Z=\{0\}$ iff both $\CL\otimes L$ and $\CL\otimes L^{-1}$ fail to be
regular.
\end{theorem}

These theorems will be proved in \S\S\ref{kine}--\ref{motion}
below. For greater clarity we describe in \S\ref{kine} the (1:1)
correspondence between triples $(A,B,D)$ (modulo conjugation) and
pairs $(S,\CL)$ where $S \subset \IP_1\times \IP_1$ and $\CL$ is a
regular line-bundle on $S$.  With this established, we show that the
DN evolution corresponds to straight-line motion on the Jacobian in
\S\ref{motion}.

Of course the correspondence between triples of matrices (or more
generally matrix polynomials) and pairs $(S,\CL)$ is a fundamental
part of the modern theory of completely integrable systems, but in
that setting the curve $S$ is naturally embedded in the total space of
$\CO(d) \to \IP_1$ (for some positive integer $d$).  The present work
may be viewed as an attempt to understand what aspects of this theory
change when $\CO(d)$ is replaced by $\IP_1\times \IP_1$.  The most
interesting feature here is the way in which a discrete-time
integrable system arises naturally from the geometry of the embedding
in $\IP_1\times \IP_1$.  From this point of view it seems natural to
ask whether the DN system is part of a discrete integrable hierarchy and
whether other interesting families of discrete integrable systems
arise by generalizations of the present construction.

\subsection{Relation to monopoles and instantons}
\label{inst}

This subsection outlines how these results were motivated by, and bear
on hyperbolic monopoles. The reader interested only in the
complete integrability of the discrete Nahm system may skip it.

When supplemented boundary conditions, there is a
correspondence, the Nahm transform, between solutions on $0 < z <2$ of
(\ref{eq1}) and solutions of the euclidean Bogomolny equations
$$
\nabla_1\Phi = [\nabla_2,\nabla_3],\;\nabla_2\Phi =
[\nabla_3,\nabla_1],\;
\nabla_3\Phi = [\nabla_1,\nabla_2],
$$
where $\nabla_j = \p_j + A_j$ are the components of a unitary
connection on $\R^3$ and $\Phi$, the Higgs field, is a section of the
adjoint bundle. On the other hand there is also a twistor
correspondence for these equations, yielding an algebraic curve $S
\subset T\IP_1$  which determines the monopole.  From the work of
Hitchin and Murray \cite{HM} one knows that the curve (\ref{eq6})
determined by the Nahm data which corresponds to a monopole, coincides
with $S$.

The discrete Nahm system arose in the work of Braam and Austin on
hyperbolic monopoles. These are solutions of the Bogomolny equations on
hyperbolic
space $H^3$, subject to certain boundary conditions \cite{MS}.  These
boundary conditions yield two numerical invariants for each solution,
the magnetic charge $k$, a positive integer, and the mass $p$, a
positive real number. When $p$ is an integer of half-integer, the
moduli space of hyperbolic monopoles of mass $p$ and charge $k$ can be
identified with a moduli space of circle-invariant instantons on $S^4$
of topological charge (instanton number) $2pk$. By decomposing the
ADHM description of such instantons under the action of the circle,
Braam and Austin proved that there is a (1:1) correspondence between
\begin{quote}
(i) Solutions of the $k\times k$ DN system in $\{1,2,\ldots,2p+1\}$,
with boundary condition that $B_1- D_1A_1$ is of rank 1, plus reality
conditions, and

(ii) hyperbolic monopoles of charge $k$ and mass $p$.
\end{quote}
(Actually Braam--Austin considered mainly the case where $2p$ is odd,
and considered their system to be defined on the set
$\{1-2p,3-2p,\ldots,-2,0,2,\ldots,2p-1\}$.)

On the other hand,
it is also known that the monopole is
determined by a spectral curve in $\IP_1 \times \IP_1$ \cite{At,MS}.
This suggested to us that the Braam--Austin system should be
integrable in terms of the geometry of such a curve.   Indeed a previous
calculation of the first author gave the equation of the spectral
curve of the monopole in terms of the corresponding Braam--Austin data
in the form
$$
\det(\eta\z\beta_i - \eta(\gamma_{i-1}^*\gamma_{i-1} +
\beta^*_i\beta_i) + \z - \beta^*_i) =0.
$$
The reader will recognize this as the equation (\ref{eq2158}) that we have
already used to associate a spectral curve to DN data.  It follows
that given a hyperbolic monopole, the associated spectral curve
$S_{mon}$, say,
agrees with the spectral curve $S_{DN}$ of the corresponding
solution of the Braam--Austin equations.  After Theorems~\ref{th1} and
\ref{th2}, it follows that there exists a holomorphic line-bundle
$\CL$ over $S_{mon}$ which gives rise to the corresponding solution of the
Braam--Austin equations. In fact we have the following
\begin{theorem}
\label{th:BA}
 Let $(A,\Phi)$ be an $SU_2$-monopole on $H^3$ with
charge $k$ and mass $p \in {\frac{1}{2}}\Z_{>0}$. Assume the spectral
curve of $(A, \Phi)$ is smooth.  Then the
solution of the DN system given by Theorem~\ref{th2}, applied to
$(S_{mon},\CO(k-1,0))$ coincides with the solution associated by Braam and
Austin to $(A,\Phi)$.
\end{theorem}

There is a slight abuse of notation here in that $\CO(k-1,0)$ is not
regular in the sense of Definition~\ref{reg}.
However, as we shall
see, $L^r(k-1,0)$ is regular for $r=1,\ldots,2p$, and the Theorem
states that the corresponding solution of DN on $\{1,\ldots,2p\}$
agrees with the one obtained from the Braam--Austin correspondence.

 This proves one of
the main parts of the following
\begin{conjecture} Let $S$ be a curve of bidegree $(k,k)$ in
$\IP_1\times \IP_1$. Then $S$ is the spectral curve of a hyperbolic
monopole of charge $k$ and  mass $p \in {\frac{1}{2}}\Z_{>0}$
iff
\begin{quote}
(o) $S$ does not intersect the anti-diagonal;

(i) $S$ has no multiple components;

(ii) $S$ is real, $L^{2p+k}_{|S}$ is holomorphically trivial;

(iii) $L^{p+1/2}(k-1,0)_{|S}$ has a real structure;

(iv) $H^0(S,L^r(k-2,0))=0$ for $r=1,2,\ldots,2p+1$.
\end{quote} \label{conj1}
\end{conjecture}

To interpret the word `real' here, recall that when $\IP_1\times\IP_1$
is viewed as the twistor space of $H^3$, it is equipped with a natural
real structure $\sigma:(p,q) \mapsto (\sigma_0(q),\sigma_0(p))$, where
$\sigma_0$ is the antipodal map. The subgroup of $SL_2(\C)\times
SL_2(\C)$ that commutes with $\sigma$ is an `antidiagonal' copy of
$SL_2(\C)$ which corresponds to the isometry group of $H^3$. This has
two orbits on $\IP_1\times \IP_1$: the anti-diagonal $\bar \triangle$,
which
is the set of all pairs $(p,\sigma_0(p))$ in $Q$, and its complement.
Condition (o) is equivalent to $S$ being a compact subset of $Q - \bar
\triangle$.  These conditions should be
compared with those for the euclidean monopole in
 \cite[p.\ 146]{NJH1}.

Of these conditions, (o) and (ii) are known from \cite{At,MS}. What is new here
is that (iii) and (iv) are proved to be necessary conditions, (iv)
being a restatement of the regularity of $L^r(k-1,0)$. Condition (iii)
follows from the reality conditions of Braam and Austin, and is
equivalent to the existence of a canonical real structure switching
the summands in
$$
H^0(S,L^{p+1/2 + t}(k-1,0))\oplus H^0(S,L^{p+1/2 -t}(k-1,0)).
$$

It is hoped
that the methods of this paper might be refined to prove that
(i)--(iv) are indeed sufficient conditions; for this one would need to
show that the solution of the DN system given by Theorem~\ref{th2}
(applied to $(S,\CO(k-1,0))$) satisfies the boundary conditions and
reality conditions written down by Braam and Austin.

A further extension of the present work might uncover the Nahm
description of {\em non-integral} hyperbolic monopoles.  Since such
monopoles still have spectral curves in $\IP_1\times
\IP_1$, it is tempting to believe that one of the corresponding
solutions of the DN equations, now defined, presumably,  on an
infinite set $Z$, should provide such a Nahm description.

Two final remarks. First, the continuum limit can be seen as the limit
as the
curvature of $H^3$ goes to $0$. In terms of the twistor spaces, this
corresponds to the singular limit of a family of embedded quadrics in
$\IP_3$.  Second, since the hyperbolic monopoles described by the
Braam-Austin system correspond to
$S^1$-invariant instantons over $S^4$,
the present work provides, in principle at least, some non-trivial solutions
to the ADHM equations\footnote{These are a system of quadratic matrix
equations}---solutions associated with algebraic curves.

\section{DN triples and spectral curves}
\label{kine}
\subsection{Notation}

From now on, we write $Q = \IP_1\times \IP_1$, with homogeneous
coordinates $[w_0,w_1]$, $[z_0,z_1]$ on the two factors. It is
sometimes helpful to think of $Q$ as $\IP(E^+)\times \IP(E^-)$, where
$E^{\pm}$ are complex symplectic vector spaces of dimension $2$, not
canonically isomorphic\footnote{The natural action of ${\rm SL}_2(\C)$
on $Q$ entails that $E^+$ and $E^-$ should be the two inequivalent
two-dimensional
irreducible representations of ${\rm SL}_2(\C)$; this corresponds to
the anti-diagonal embedding mentioned above}. The symplectic form
in $E^{\pm}$ will be denoted $\langle\cdot,\cdot\rangle$.

In particular, then, if $\CO(a,b)=
p_1^*\CO(a)\otimes p_2^*\CO(b)$, there are canonical isomorphisms
\be{eq211}
H^0(Q,\CO(1,1)) = E^+\otimes E^-,\;\;H^0(Q,\CO(a,b)) = S^a E^+\otimes
S^b E^-
\end{equation}
provided $a$ and $b$ are non-negative. (Here $S^m$ denotes $m$-th
symmetric power.) We shall usually denote by $L$
the line-bundle $\CO(1,-1)$.

The first of (\ref{eq211}) leads to an evaluation map $E^+\otimes
E^-\otimes \CO
\to \CO(1,1)$ whose kernel $\CK$ will be very important in what
follows.  In addition to the defining exact sequence
\be{eq212}
0 \to \CK \to E^+\otimes E^-\otimes \CO \to \CO(1,1) \to 0,
\end{equation}
(where the second map is given by $g_1\otimes g_2 \mapsto \langle
g_1,w\rangle\langle g_2,z\rangle$)
we have an exact sequence
\be{eq213}
0 \to \CO(-1,-1) \to E^+\otimes\CO(0,-1) \oplus E^-\otimes\CO(-1,0)
\to \CK \to 0.
\end{equation}
Here the maps are $f \mapsto (f\otimes w,f\otimes z)$ and $(g_1,g_2)
\mapsto g_1\otimes z - g_2\otimes w$, $w$ and $z$ standing for the
tautological sections of $E^+(1,0)$ and $E^-(0,1)$ respectively.

If a local section of $\CK$ is represented in the form $g_1\otimes z -
g_2\otimes w$ then $\langle g_1,w\rangle$ and $\langle g_2,z\rangle$
are local sections of $L$ and $L^{-1}$ respectively.
Since each of these vanish if $(g_1,g_2)$ is in the image
of $\CO(-1,-1)$ in (\ref{eq213}), we obtain sheaf maps $\CK \to L^{\pm
1}$ and hence exact sequences
\be{eq214}
0 \to E^-\otimes \CO(-1,0) \to \CK \to L \to 0
\end{equation}
and
\be{eq215}
0 \to E^+\otimes \CO(0,-1) \to \CK \to L^{-1} \to 0.
\end{equation}
These four exact sequences will be much used below.

\subsection{DN maps and spectral curves}

In order to give an invariant statement of the correspondence between
matrix data and holomorphic line-bundles over algebraic curves, we
introduce the notion of a {\em DN map of charge} $k$. This is just an
{\em injective} complex-linear map $\alpha: \C^k \to \C^k\otimes E^+
\otimes E^-$. Choosing a basis in $E^+\otimes E^-$ we may consider
$\alpha$ as a list $(\alpha_{00},\alpha_{01},\alpha_{10},\alpha_{11})$
of maps $\C^k \to \C^k$. It will turn out to be natural to regard the
two copies of $\C^k$ as different, so that the natural notion of
equivalence is given by the action of $GL_k\times GL_k$ on $\alpha$,
$(g_1,g_2)\alpha = g_1\alpha g_2^{-1}$. In particular any injective
linear map $\alpha:U\to V\otimes E^+\otimes E^-$, where $\dim U = \dim
V =k$ gives rise to a DN map, by choosing bases in $U$ and $V$. The
freedom in choosing these bases corresponds precisely to the
$GL_k\times GL_k$-action just mentioned.  In our application,
$\alpha_{10}$ is invertible and may be used to identify the two copies
of $\C^k$. Then the DN map takes the form $(A,B,1,D)$ and $A$, $B$ and
$D$ will be identified with the matrix data in \S\ref{sr}.

The {\em spectral curve} $S(\alpha)$ associated to a DN map $\alpha$
is defined as follows. Identifying $E^+$ with the space of sections of
$\CO(1,0)$ and $E^-$ with the space of sections of $\CO(0,1)$, we may
think of $\alpha$ as an element of $H^0(Q,\C^k\otimes\C^k\otimes\CO(1,1))$.
Because $\alpha$ is assumed injective, taking the determinant, we
obtain $0\not=\det\alpha \in H^0(Q,\CO(k,k))$. Then we put
\be{eq221}
S(\alpha):= \{ \det\alpha =0\}.
\end{equation}
Thus $S(\alpha)$ is an algebraic curve of bidegree $(k,k)$ in $Q$.
In addition, we define a sheaf $\CL(\alpha)$ by the
exactness of
\be{eq224}
0\to \C^k\otimes \CO(-1,-1) \stackrel{\alpha}{\longto} \C^k \to
\CL(\alpha) \to 0.
\end{equation}
We shall refer to $(S(\alpha),\CL(\alpha))$ as the {\em spectral data}
determined by $\alpha$.  It is clear that the spectral data depends
only on the $GL_k\times GL_k$-equivalence class of $\alpha$ and that
the support of $\CL(\alpha)$ is contained in $S(\alpha)$.

Twisting (\ref{eq221}) by $\CO(0,-1)$ and $\CO(-1,0)$ and taking the
corresponding long exact sequences, we note that
$$
H^i(\CL(\alpha)(-1,0)) = 0,\;\;
H^i(\CL(\alpha)(0,-1)) = 0 \mbox{ for all }i.
$$
Since the genus of $S$ is $(k-1)^2$, if $E$ is a bundle over $Q$ of
rank $n$ and bidegree $(a,b)$,
\be{eq22151}
\ind(E):= \dim H^0(S,E) - \dim H^1(S,E) = k(a+b) -nk(k-2)
\end{equation}
(Riemann--Roch).
In particular, the above vanishing of cohomology implies that the
degree of $\CL(\alpha)(-1,0)$ is $k(k-2)$; so the degree of $\CL$ is
$k(k-1)$. Accordingly, let $J=J(S)$ denote the set of
holomorphic line-bundles on $S$ of degree $k(k-1)$.

\begin{definition}\label{reg} The element $\CL \in J$ is called {\em regular}
iff both $\CL(-1,0)$ and $\CL(0,-1)$ are in the complement of the
$\vartheta$-divisor; i.e.\ if and only if
\be{eq222}
H^0(S,\CL(-1,0)) =  H^1(S,\CL(-1,0)) = 0,\;
H^0(S,\CL(0,-1)) =  H^1(S,\CL(0,-1))=0.
\end{equation}
The set of regular elements of $J$ is denoted by $J^{reg}$.
\end{definition}

The `kinematic' part of our construction now has the following
statement:
\begin{theorem} Let $S$ be a smooth curve in $Q$ of bidegree
$(k,k)$. Then there
is a natural bijection between $J(S)^{reg}$ and the set
$$
\{\alpha : \C^k \to \C^k\otimes E^+\otimes E^-|\,\,S(\alpha) =
S\}/GL_k\times GL_k.
$$
\label{kinc}\end{theorem}

\noindent{\bf Proof}  In one direction this bijection is the map which
assigns to $\alpha$ the spectral data $(S(\alpha),\CL(\alpha))$.  We
shall show that $\CL(\alpha)$ is a holomorphic line-bundle on
$S$, i.e.\
a locally free sheaf of $\CO_S$-modules
of rank 1.
To prove that $\CL(\alpha)$ is a sheaf of $\CO_S$ modules it is necessary and
sufficient to show that the ideal $\CI \subset \CO$ defined by
$\det\alpha$ annihilates $\CL(\alpha)$. In other
words, if $v\in \C^k$ then $(\det\alpha)v$ is in
the image of $\alpha$. But by definition, the matrix $\beta$ of cofactors of
$\alpha$ satisfies $\alpha\beta = \beta\alpha = \det\alpha\cdot {\rm
Id}$. Hence $\alpha$ carries $\beta v$ to
$(\det\alpha)v$, as required.

We now use the assumption that
$S$ is smooth to prove that $\CL(\alpha)$ has rank
$1$. For suppose not. Then there is a point $(\eta_0,\zeta_0)$ on
$S$ such that  the nullity $n$ of $\alpha(\eta_0,\zeta_0)$ is at least 2. By
replacing $\eta$ by $\eta-\eta_0$ and $\zeta$ by $\zeta -\zeta_0$ we
may suppose this point is $(0,0)$. With such a choice of coordinates,
$\alpha$ takes the form
\be{eq225}
\alpha(\eta,\zeta) = \eta\zeta \alpha_{00} + \eta \alpha_{01} + \zeta
\alpha_{10}  + \alpha_{11}
\end{equation}
where the nullity of $\alpha_{11}$ is equal to $n$. Choosing an appropriate
basis of $V$, we may suppose that the first two rows of $\alpha_{11}$ are
identically zero. But now when we expand $\det \alpha(\eta,\zeta) = a\eta
+ b\zeta + \ldots$ in ascending powers of $(\eta,\zeta)$, we have
$a=b=0$. For each term in the expansion of the determinant  contains
an entry from the first row of $\alpha(\eta,\zeta)$ and an entry from
the second row, so each
term has a factor of $\eta^2$, $\eta\zeta$, or $\zeta^2$. Hence the
curve is singular at $(0,0)$, contradiction.

Finally let us show that $\CL(\alpha)$ is locally free. This is a local
question, so we may once again assume that we are working near the
point $(0,0)\in S$. Choose the basis in $\C^k$ so that the first
row of $\alpha_{11}$ in (\ref{eq225}) is identically zero, the
remaining rows being linearly independent.  Then obviously $\CL(\alpha)_0$ is
generated by the first basis vector $e_1$, but by continuity, the same
is true of $\CL(\alpha)_x$ for all $x\in S$ sufficiently close to $0$. In
other words there exists an open set $U$ of $S$ containing $0$ such
that multiplication by $e_1$ followed by projection to $\CL(\alpha)$ gives an
isomorphism $\O_S|U \cong \CL(\alpha)|U$, as required for
$\CL(\alpha)$ to be locally free.  Since we have already seen that
$\CL(\alpha)(-1,0)$ and $\CL(\alpha)(0,-1)$ have no cohomology this
completes the map from the set of DN maps to $J^{reg}$.

To go in the other direction, we show how to construct a DN map
$\alpha(\CL)$, given a pair $(S,\CL)$ with $\CL$ in $J^{reg}$.   Given
such $\CL$, we may consider the sheaf cohomology groups
$$
U(\CL):= H^0(S,\CK\otimes \CL)\mbox{ and }V(\CL):= H^0(S,\CL).
$$
By definition of $\CK$ there is a natural map $\alpha(\CL):
U \to V\otimes E^+\otimes E^-$.  The next three lemmas are devoted to
showing that $\alpha(\CL)$ is a DN map with spectral curve equal to
$S$.  Namely we establish in turn that $U(\CL)$ and $V(\CL)$ are of
the correct dimension $k$, that $\alpha(\CL)$ is injective, and that
$S= S(\alpha(\CL))$.

\begin{lemma} If $\CL\in J(S)^{reg}$ then $H^0(S,\CL)$,
$H^0(S,\CL\otimes L)$ and $H^0(S,\CL\otimes L^{-1})$ are
$k$-dimensional.
\label{kdim}\end{lemma}

\noindent{\bf Proof} Let $C^+$ be a generator of $Q$ in the linear
system of $\CO(1,0)$. Then we have the structure sequence
\be{eq226}
0 \to \CO(-1,0) \to \CO \to \CO_{C^+} \to 0.
\end{equation}
Since $S$ is smooth, we may choose $C^+$ so that $S\cap C^+$ consists of
$k$ distinct points. Then we obtain
\be{eq227}
0 \to \CO_S(-1,0) \to \CO_S \to \CO_{S\cap C^+}\to 0
\end{equation}
and the latter is a skyscraper sheaf, supported at the $k$ points of
$S\cap C^+$. Tensor (\ref{eq227}) with $\CL$ and take global sections,
to obtain
$$
\cdots\to H^0(S,\CL(-1,0)) \to H^0(S,\CL) \to \C^k \to H^1(S,\CL(-1,0)
\to\cdots.
$$
Hence if $\CL$ is regular, $H^0(S,\CL)$ is $k$-dimensional, by
evaluation of sections on  a generically chosen generator
$C^+$. The same argument works for $H^0(S,\CL\otimes L)$, for
$\CL\otimes L(-1,0)= \CL(0,-1)$ has no cohomology.

Similarly, evaluation of sections on a generic generator $C^-$
in the linear system of $\CO(0,1)$ shows that $H^0(S,\CL)$ and
$H^0(S,\CL\otimes L^{-1})$ are both $k$-dimensional. QED

\begin{lemma}
\label{lemma:fibre} If $\CL$ is regular, then $H^0(S,\CK\otimes \CL)$ is
$k$-dimensional, and the sequence
$$
0 \to H^0(S,\CK\otimes \CL) \stackrel{\alpha}{\longrightarrow}
H^0(S,\CL)\otimes E^+\otimes E^-
\stackrel{m}{\longrightarrow} H^0(S,\CL(1,1)) \to  0
$$
is exact.  Here $m$ is the multiplication map which arises by identifying
$E^+\otimes E^-$ with $H^0(Q,\CO(1,1))$.
\label{mmap}\end{lemma}

\noindent{\bf Proof} It is plain that the sequence in question arises
by tensoring (\ref{eq212}) with $\CL$ and taking global
sections. Thus $m$ is surjective if $H^1(S,\CK\otimes \CL)$
vanishes. Tensoring (\ref{eq213}) with $\CL$ and taking the long exact
sequence yields a surjective map
$$
E^+\otimes H^0(S,\CL(0,-1)) \oplus E^-\otimes H^0(S,\CL(-1,0)) \to
H^1(S,\CK\otimes \CL),
$$
so that if $\CL$ is regular, $H^1(S,\CK\otimes\CL)=0$.

On the other hand, from (\ref{eq212}), $\CK$ is of rank $3$ and bidegree
$(-1,-1)$, $\CK(a,b)$ has bidegree $(3a-1,3b-1)$ and from (\ref{eq22151}),
$$
\ind(\CK(a,b)) = k(3a+3b-2) -3k(k-2) = k(3(a+b-k) + 4).
$$
Hence $\ind(\CK\otimes \CL) = k$ and so $H^0(S,\CK\otimes\CL)$ is
$k$-dimensional. QED

To summarise: given a DN-map $\alpha$ we can construct a pair
$(S(\alpha), \CL(\alpha))$ consisting of a
 smooth curve $S$  in $Q$ of bi-degree $(k,k)$
and  a regular line bundle $\CL \to S$; conversely  from such a pair $(S,
\CL)$
we can construct a DN-map $\alpha(\CL)$.  Obviously we would like
these two constructions to invert each other. We will  show this by
showing that
if we start with a pair $(S, \CL)$ then the spectral curve and
line bundle, say $(S', \CL')$ constructed from the DN-map
of $\alpha(\CL)$ is isomorphic to $(S, \CL)$.

Consider the sequence
$$
H^0(S, \CK \otimes \CL) \otimes \O_Q(-1, -1)
\overset{M}{\to} H^0(S, \CL)\otimes \O_Q \overset{\text{ev}}{\to}
\O_S(\CL)
$$
Then  $S'$ is defined by  $\det(M) = 0$ and $\CL'$ is the cokernel of
$M$.

Clearly  $\text{ev}\circ M = 0$,
moreover the proof of Lemma \ref{kdim}
shows that for generic points $z$ of $S$ we can
always find a section $\psi$ of $\CL$ with $\psi(z) \neq 0$.
At these points  the kernel
of $\text{ev}$ has at least co-dimension $1$ and thus
 $\det(M) $ vanishes.   From this we conclude
that  $\det(M) = 0$ on $S$ so that $S \subset S'$ but both
are smooth curves of the same bidegree so they
must be equal.
The cokernel of $M$ is now a line bundle
over $S= S'$.  Because  $\text{ev}$ vanishes
on the image of $M$ it induces a map from
$\text{coker} M = \CL'$ to $\CL$.  As both of these are line
bundles of the same degree this  map is a
holomorphic section of the trival bundle $\CL'\otimes \CL^*$.
and hence either the zero section or everywhere non-vanishing.
However  this section  is non-vanishing
at points of $S$ for which there is a non-vanishing section of $\CL$.
 As we have already argued this
happens generically on $S$ and thus $\CL$ and $\CL'$
are isomorphic.

\section{DN evolution and motion on the Jacobian}
\label{motion}
We turn now to an explanation of the claim that the discrete-time
evolution of the
DN equations corresponds to straight-line motion on $J(S)$. More
precisely we shall prove Theorems~\ref{th1} and \ref{th2} here.

\subsection{Solutions of the DN system from an algebraic curve}

In this section we shall prove Theorem~\ref{th2}. So we assume given a
smooth curve $S$ of bidegree $(k,k)$ in $Q$ and a regular line bundle
$\CL$ over $S$.  By moving $S$ by an element of ${\rm SL}_2(\C)$, we
may suppose that the point with coordinates $((0:1),(1:0))$ does {\em
not} lie on $S$.

According to Theorem~\ref{kinc}, these data give a DN map
$$
\alpha = \alpha_{00}w_0z_0 + \alpha_{01}w_0z_1 +
\alpha_{10}w_1z_0 + \alpha_{11}w_1z_1$$
such that each of the $\alpha_{ij}$ maps $U(\CL) =
H^0(S,\CK\otimes\CL)$ into $V(\CL) = H^0(S,\CL)$.  Evaluating at
$((0:1),(1:0))$ gives the element $\alpha_{01}$; since this is not on
$S$, it follows that $\alpha_{01}$ is an isomorphism.  Thus we may
break the symmetry of the problem by using $\alpha_{01}$ to identify
$U(\CL)$ with $V(\CL)$.  Having done so, the DN map takes the form
$$
M(w,z) = w_0z_0\, A + w_0z_1\, B + w_1z_0 + w_1z_1\, D
$$
where $A$, $B$ and $D$ are endomorphisms of a $k$-dimensional vector
space.  These will be identified with the DN data of the same name
that were introduced in \S\ref{sr}.

In order to complete the definition of DN data, we must define the
operators $P^{\pm}$. These arise directly from the geometry of
$\IP_1\times \IP_1$ from the basic exact sequences
(\ref{eq214}) and (\ref{eq215}) as follows. After tensoring with
$\CL$, we obtain from the corresponding long exact sequences,
\be{eq231}
\cdots E^-\otimes H^0(S,\CL(-1,0)) \to U(\CL)
\stackrel{\phi^+}{\longto} V(\CL\otimes L) \to
E^-\otimes H^1(S,\CL(-1,0)) \cdots
\end{equation}
and
\be{eq232}
\cdots E^+\otimes H^0(S,\CL(0,-1)) \to U(\CL)
\stackrel{\phi^{-}}{\longto} V(\CL\otimes L^{-1}) \to
E^+\otimes H^1(S,\CL(0,-1)) \cdots.
\end{equation}
Since $\CL$ is assumed regular, $\phi^{\pm}$ are isomorphisms.  On the
other hand we have just identified $U(\CL)$ with $V(\CL)$, so
$\phi^{\pm}$ gives rise to an isomorphism
$P^{\pm}: V(\CL) \to V(\CL\otimes L^{\pm 1})$.

Now suppose that $\CL$ and $\CL\otimes L$ are both regular. Then the
construction we have just described yields $k$-dimensional vector
spaces $V_0 = V(\CL)$, $V_1 = V(\CL\otimes L)$, maps $M_0$, $M_1$,
where
\begin{equation} \label{mdef}
M_r(\eta,\zeta) = \eta\zeta\, A_r + \eta\, B_r + \zeta + D_r
\end{equation}
and operators $P^+_0: V_0 \to V_1$, $P^-_1: V_1 \to V_0$.  (Here
$\eta= w_0/w_1,\zeta = z_0/z_1$ are being used to make the
formulae more readable.)  Unravelling the definitions, we obtain the following
formulae relating these maps:
\begin{eqnarray}
M_r(\eta,\zeta)s_r(\eta,\zeta)  & = &0, \\
(\eta A_0 + 1)s_0(\eta,\zeta) &=&  P_0^+s_0(\eta,\zeta), \label{zpp} \\
(\eta B_0 + D_0)s_0(\eta,\zeta) &=& -\zeta P_0^+s_0(\eta,\zeta), \label{pp} \\
(\zeta A_1 + B_1)s_1(\eta,\zeta) &=&  P_1^-s_1(\eta,\zeta), \label{epm} \\
(\zeta + D_1)s_1(\eta,\zeta) &=& - \eta P_1^-s_1(\eta,\zeta),
\label{pm} \\
&&\mbox{for all } s_r \in V_r \mbox{ and }(\eta,\zeta) \in S \nonumber
\end{eqnarray}
The equations have been written at such length to emphasize that here
$(\eta,\zeta)$ are not independent parameters; they live on $S$.

\begin{proposition} The data defined by equations
(\ref{mdef}--\ref{pm}) satisfy the DN equations on $[0,1]$:
$$
P^+_0D_0 = D_1P^+_0,\;\;P_1^-P_0^+ + A_0D_0 = B_0,\;\;
P^-_1 A_1 = A_{0}P^-_{1},\;\;P_1^-P_{0}^+ + D_1A_1 = B_1.
$$
\end{proposition}

\noindent{\bf Proof}: Since $P^+_0$ is a map $V_0 \to
V_1$, the definition of $M_1$ implies that
\begin{equation} \label{d1}
[\eta\zeta A_1P_0^+ + \eta B_1P_0^+  + \zeta P_0^+ +
D_1P_0^+]s_0(\eta,\zeta) = 0.
\end{equation}
Use (\ref{epm}) to combine the first two terms, and (\ref{zpp}) in
the third to obtain
\begin{equation} \label{d2}
[\eta P^-P^+ -\eta B_0 - D_0            +D_1P^+_0]s_0(\eta,\zeta)= 0.
\end{equation}
Now write
\begin{equation} \label{d3}
DP^+ = [D,P^+] + P^+D = [D,P^+] + (1 + \eta A_0)D_0
\end{equation}
making use of (\ref{zpp}). Substituting this into (\ref{d2}) we
obtain, finally
\begin{equation} \label{d4}
[\eta(P_1^-P^+_0 - B_0 + A_0D_0) + [D,P^+]]s_0(\eta,\zeta) = 0.
\end{equation}
One derives similarly from $M_0P^-_{1}=0$ a linear combination of the
other two equations.  The proof is now completed with the aid of the
lemma below which says that we can conclude from (\ref{d4}) that the
two terms must vanish separately. QED

\begin{lemma}  Suppose the relation $(Q_0\eta + Q_1)s(\eta,\zeta)$
holds for matrices $Q_0$ and $Q_1$ and all $(\eta,\zeta) \in S$, $s\in
V(\CL)$. Then $Q_0=0$, $Q_1=0$.
\end{lemma}

\noindent{\bf Proof} If not there exist $a\in\C$, $s\in V(\CL)$, such
that $(Q_0a+Q_1)s\not=0$, where we may suppose $a$ has the property that
the intersection of $S$ with $\{\eta=a\}$ consists of $k$ distinct
points $(a,b_1),\ldots,(a,b_k)$. From the given relation, we
have
$$
(Q_0a + Q_1)s(a,b_j) = 0\mbox{ }j=1,\ldots,k.
$$
Because $\CL$ is regular,
$V(\CL)$ is identified with $\C^k$ by evaluation at these points
(Lemma~\ref{kdim}). This
contradicts the assumption that $(Q_0a+Q_1)s\not=0$. QED

\subsection{From the DN equations to spectral data}
\label{LAX}
We now turn to the proof of Theorem~\ref{th1}.
The basic idea here has already been described in \S\ref{sr} and at
greater length in Theorem~\ref{kinc}. What remains to be proved is
that if we have a solution of the DN system in $[0,1]$ (say), then the
two spectral curves $S_0$ and $S_1$ coincide, and that the two
line-bundles $\CL_0$ and $\CL_1$ satisfy $\CL_1 = \CL_0\otimes L$.

For this a Lax formulation of the DN system is
needed; for this we are indebted to Richard Ward \cite{W}, who noted that
$$
  {\hat W}^+ = P^+ - \lambda A
$$
and
$$
{\hat W}^- =  P^- + \lambda^{-1}D
$$
form a Lax pair for the discrete Nahm equations. In order to interpret
these formulae it is essential to think in terms of `discrete gauge
theory', as follows.

Given $Z = \{r_0,\ldots, r_1\}$ we may think of the vector spaces $V_r$
as forming a vector bundle $\V$ over the discrete space $Z$.  The $A$,
$B$, $D$ become sections of the corresponding bundle of endomorphisms,
while $P^\pm$ are the discrete analogue of connections (more
precisely, of parallel transport operators). Denote by $\Ga(\V)$ the
space of sections of $\V$; this is just the set of sequences $f_r$
with $f_r\in V_r$ for all $r$.  Then the formulae for ${\hat W}^\pm$
make sense as operators on $\Ga(\V)$. Specifically, if $f\in \Ga(\V)$,
then
$$
(\hat{W}^+f)_r = P^+_{r-1}f_{r-1} - \lambda A_r f_r.
$$
Ward's observation is that the condition $[{\hat W}^+, {\hat W}^-]
=0$, for all values of $\lambda$, is equivalent to the DN equations.

To recover $S$, we follow standard practice and ask for simultaneous
eigensections for ${\hat W}^{\pm}$.  The commutativity of ${\hat W}^+$
and ${\hat W}^-$
means that ${\hat W}^+$ acts on any eigenspace of ${\hat W}^-$. The
conditions that
${\hat W}^{\pm}$ have simultaneous eigensections defines an algebraic curve;
in the right coordinates, this curve is given precisely by $\det
M(\eta,\zeta) = 0$.

In order to put this plan into action, we shall replace the above
operators by
$$
W^+ = P^+ - \eta A - 1
$$
and
$$
W^- =\eta P^- + \zeta + D
$$
and study the conditions on $(\eta,\zeta)$ under which there exists a
section $f$ which satisfies
$$
W^+f = 0\;\;\; W^-f = 0.
$$
Of course Ward's original operators are recovered by deleting $1$ from
the definition of $W^+$ and $\zeta$ from the definition of $W^-$.
This particular modification is motivated by the definitions
(\ref{zpp}) and (\ref{pm}) of $P^{\pm}$ above.  For future
reference note that in homogeneous coordinates, the Ward operators become
\begin{equation} \label{eq309}
W^+ = z_1P^+ - w_0 A -w_1,\;\;\;W^- = w_0P^- + z_0 + z_1D.
\end{equation}

Let $K^{\pm}= K^{\pm}(\eta,\zeta)$ be the space of $W^{\pm}$-parallel
sections of $\V$.  If the rank of $\V$ is $k$ then since a parallel
section is determined by its value at any point, $K^{\pm}$ is a
$k$-dimensional complex vector space and can be identified with any
one of the $V_r$ (by evaluation). We shall look for the condition on
$(\eta,\zeta)$ that makes the intersection $K^+\cap K^-$ inside
$\Gamma(\V)$ non-trivial.
\begin{theorem} Given data $A$, $D$, $P^\pm$ satisfying the discrete
Nahm equations, let
$$
B = P^+P^- + DA = P^-P^+ + AD
$$
and consider
$$
M = \eta\zeta A + \eta B + \zeta +D.
$$
Then the condition $\det M = 0$, viewed as an equation for
$(\eta,\zeta)$, is independent of $r$ and is equivalent to the
condition that $K^-(\eta,\zeta)\cap K^+(\eta,\zeta) \not= 0$.
\end{theorem}

\noindent {\bf Proof}: Using the above formulae,
$$
M =  \eta\zeta A + \eta (P^-P^+  +AD) + \zeta + D
  = \eta\zeta A + \eta P^-(W^+ + \eta A + 1) + \eta AD + \zeta  + D
$$
$$
\;\;\;\;\;\;\;\;\;\;
= \eta P^- W^+ + (\eta P^- + \zeta + D) + \eta A(\eta P^- + \zeta + D)
$$
where we have used $[P^-,A]=0$. Recognizing $W^-$ in the second and
third terms of this we obtain
$$
M = \eta P^- W^+ + (\eta A + 1)W^- = \eta P^-W^+ + P^+ W^- - W^+W^-.
$$
Because $[W^+,W^-]=0$, this also yields
$$
M = \eta P^- W^+ + P^+ W^- - W^-W^+.
$$
In particular we see that
$$
M|K^-  = \eta P^-W^+,\;\;\; M|K^+ = P^+W^-
$$
It follows that $K^+ \cap K^- \not= 0$ iff $\det M =0$.  Since the
first condition is independent of $r$, it follows that the condition
$\det M =0$ is also.  The key point is that the operator $M$ on
$\Gamma(\V)$ is of order zero in the sense that its value at $V_r$
depends only upon $f_r$. QED

In order to derive the relation between $\CL_1$ and $\CL_0$, it is
convenient to dualize. Then given $g_r^t\in V^*_r$, we may consider
\be{eq311}
[g^tW^+]_0 = g^t_1 P^+_0 - g_0^t(\eta\, A_0 + 1)
\end{equation}
and
\be{eq312}
[g^tW^-]_1 = \eta\,g^t_0 P^-_1 + g_1^t(\zeta + D_1).
\end{equation}
\begin{proposition} Let $(A,B,D,P^{pm})$ satisfy the DN equations in
$[0,1]$.
Let $g_r^t \in V_r^*$ ($r=0,1$) satisfy
$[g^tW^-]_1= 0$. Then
\begin{quote}
(a) If $g_1^tM_1 =0$  we have also $[g^tW^+]_0 = 0$ and $g^t_0M_0= 0$

(b) If $g_0^tM_0 =0$  we have also $[g^tW^+]_0 = 0$ and $g^t_1M_0= 1$
\end{quote}
\end{proposition}

Before giving the proof note that dualizing the sequence which defines
$\CL_r$, we obtain on $S$
$$
0\to \CL_r^* \to V_r^* \stackrel{M^t_r}{\longrightarrow} V_r^*(1,1)
$$
and that according to this lemma $W^-$ defines an identification
$\CL_0^*\otimes L^{-1} \to \CL_1^*$ (recall the homogeneity
(\ref{eq309}) of $W^-$). Hence this result completes the proof
of Theorem~\ref{th1}.

\noindent{\bf Proof of Proposition} We shall only do part (a), since part
(b) is very similar. Since $B_1 = (P_0^+P^-_1)+ D_1A_1$,
we have
\begin{equation} \label{og2}
g_1^tM_1 = g_1^tP^+(\eta P^-) + g_1(\zeta + D_1)(1 + \eta A_1).
\end{equation}
But we are given
\begin{equation} \label{og3}
g_1^t(\zeta + D_1) = - \eta\,g_0^tP^-_1
\end{equation}
so inserting this in (\ref{og2}),
$$
g_1^tM_1 = - g_1^tP_0^+(\eta P_1^-) + g_0^t(1+\eta A_0)\eta P^-_1,
$$
where we have used the DN equation $A_0P_1^- = P_1^-A_1$. But the RHS
is now equal to $-\eta\,[g^tW^+]_0 P^-_1$. This proves the first part
of (a).

We now have the following three equations:
\begin{eqnarray}
g_1^t(\eta\zeta\, A_1 + \eta\, B_1 + \zeta + D_1) & = & 0\\
g_0^t(\eta\,P_1^-) & = &- g_1^t(\zeta + D_1) \\
g_1^t P_0^+& =&  g_0^t(\eta A_0 + 1) \\
\end{eqnarray}
From the first of these, there exists a vector $h$ such that
$$
h^t(1,-\eta) = g_1^t(\zeta A_1 + B_1, \zeta + D_1)
$$
and comparing with the second, $h^t = g_0^tP_1^-$, so
$$
g_0^tP_1^- = g_1^t(\zeta A_1 + B_1).
$$
Now expand $g_0^tM_0$ using $B_0 = P^-P^+ + A_0D_0$
\begin{equation} \label{og1}
g_0^tM_0 = g_0^t\eta P_1^-P_0^+ + g_0^t(1 + \eta A_0)(\zeta + D_0)
=  \eta\,g_1^t(\zeta A_1 + B_1)P^+ - g_1^tP_0^+(\zeta + D_0).
\end{equation}
Since $P_0^+D_0 = D_1 P_0^+$, we may rearrange this to obtain
$$
g_0^tM_0 =  g_1^tM_1 P^+,
$$
completing the proof of (a). QED

\section{Application to hyperbolic monopoles}

We now want to relate the general integration of the DN system to the
particular solutions that correspond to hyperbolic monopoles, as
discussed in \S\ref{inst} and prove Theorem \ref{th:BA}.

We start by assuming that
$S$ is the spectral curve of
a hyperbolic monopole of charge $k$ and mass $p\in
{\frac{1}{2}}\Z$. We wish to prove first that
$L^r(k-1,0)$ is regular for $r=1,\ldots,2p$, i.e. the vanishing theorem
$$
V_r=H^0(S,L^r(k-2,0) = 0\mbox{ if }r=1,2,\ldots,2p+1.
$$
The first step in the proof is to interpret each of these groups in
terms of the cohomology of the bundle $E \to Q$ that corresponds, by
twistor theory, to the hyperbolic monopole.  This starts from the
description of $E$ in terms of $S$, by the two exact sequences
\be{eq41}
0\to L^{-p}(-k,0) \to E \to L^p(k,0) \to 0
\end{equation}
and
\be{eq42}
0\to L^{p}(0,-k) \to E \to L^{-p}(0,k) \to 0
\end{equation}
related by the real structure. The obvious composites
$L^{-p}(-k,0) \to L^{-p}(0,k)$ and $L^{p}(0,-k) \to L^{p}(k,0)$ are
both given by multiplication by $F$, the equation of the spectral
curve $S$. It follows that $L^{2p+k}$ is holomorphically trivial over $S$.

From the structure sequence
\be{eq43}
0 \to \CO(-k,-k) \to \CO \to \CO_S \to 0
\end{equation}
we obtain the exact sequence
$$
\to H^0(Q,L^r(k-2,0)) \to H^0(S,L^r(k-2,0))
\stackrel{\delta}{\longrightarrow} H^1(Q,L^r(-2,-k)) \to
$$
and for $r$ in the given range, $H^0(Q,L^r(k-2,0)) =0$, so that
$\delta$ is injective.  Now tensor (\ref{eq42}) with $L^{r-p}(-2,0)$
to get
\be{eq44}
0\to L^{r}(-2,-k) \to EL^{r-p}(-2,0) \to L^{r-2p}(-2,k) \to 0.
\end{equation}
From this we obtain an exact sequence
$$
H^0(Q,L^{r-2p}(-2,k)) \to H^1(Q,L^r(-2,-k)) \to H^1(Q,EL^{r-p}(-2,0))
\to
$$
and again, for $r=1,\ldots,2p+1$, the first space vanishes. It follows
that the composite map
$$
H^0(S,L^{r}(k-2,0)) \to H^1(Q,EL^{r-p}(-2,0))
$$
is injective.  Next we claim that the right-hand side is a
summand in the group $H^1(\IP_3,\tE(-2))$ where $\tE$ is the bundle
which represents the $S^1$-invariant instanton corresponding to our
given hyperbolic monopole.  The non-trivial point in this is the
following. There is a map $\IP_3^0 =\IP_3 - L^+\cup L^- \to Q$ where $L^{\pm}$
are projective lines. Then $\tE|\IP_3^0 = \pi^*(E)$, but when $p$ is
integral or half-integral, this extends to a holomorphic bundle over
the whole of $\IP_3$, corresponding to the extension of the
$S^1$-invariant instanton to all of $S^4$.

In this situation, one always has $H^1(\IP_3,\tE(-2))=0$ \cite{ADHM},
so if the above claim is true, then we obtain the vanishing of $V_r$
that we require.  Since $\pi^*$ gives a map
$$
H^1(Q,EL^{r-p}(-2,0)) \to H^1(\IP_3^0,\tE(-2))
$$
the main thing is to show that everything in the image of this map
extends to $\IP_3$.  For this we must go into the description of $E$
and $\tE$ in more detail, and describe in particular the extension of
$\pi^*(E)$ to $\IP_3$.

\subsection{Background on $E$ and $\tE$}

Let us begin with the description of the map $\pi$. In terms of
homogeneous coordinates $(z_0,\ldots,z_3)$ in $\IP_3$, the
$\C^\times$-action is given by
$$
(z_0,z_1,z_2,z_3) \mapsto (\lambda^{1/2}z_0,\lambda^{-1/2}z_1,
\lambda^{1/2}z_2,\lambda^{-1/2}z_3),
$$
the fixed set consists of
$$
L^+ = \{z_0= z_2=0\} \mbox{ and }
L^- = \{z_1= z_3=0\}
$$
The map $\pi$ is the corresponding quotient map and gives affine
coordinates $\eta = z_0/z_2$, $-1/\zeta = z_1/z_3$ in $Q$. In these
coordinates, the anti-diagonal has the equation $1
+\eta\overline{\z}=0$.

Now $\pi^*\CO(a,b)$ is isomorphic to $\CO(a+b)$ over $\IP_3^0$; the
different possible values of $a$ and $b$ are distinguished on
$\IP^0_3$ by the different possible lifts of the $\C^\times$-action.
Indeed, the lifted action on $\pi^*\CO(a,b)$ is given by
$$
\pi^*f(\lambda^{1/2}z_0,\lambda^{-1/2}z_1,
\lambda^{1/2}z_2,\lambda^{-1/2}z_3) =
\lambda^{(a-b)/2}\pi^*f(z_0,z_1,z_2,z_3).
$$
Combining this with the overall homogeneity of $f$, we have found a
way to represent the pull-back of
a local section $f$ of
$\CO(a,b)$ on $\IP_3^0$: as a function satisfying
$$
\pi^*f(\lambda z_0,z_1, \lambda z_2, z_3)  = \lambda^a \pi^*
f(z_0,z_1,z_2,z_3)
$$
and
$$
\pi^*f( z_0,\lambda z_1, z_2, \lambda z_3)  = \lambda^b \pi^*
f(z_0,z_1,z_2,z_3)
$$
In particular, such a section extends smoothly through $L^+$ if $a\geq
0$ and through $L^-$ if $b\geq 0$.
Using this idea, we can see how $\pi^*E$ extends to $\IP_3$ as
follows.

The extension class defining (\ref{eq41}) is defined by taking a
trivialization $s$ of $L^{2p+k}$ and mapping it (via
(\ref{eq43})$\otimes L^{2p+k}$) into
$H^1(L^{2p+k}(-k,k))$.  Explicitly, we may take $s_+$ to be smooth
section of $L^{2p+k}$ supported near to $S$, such that $s_+\not=0$ on
$S$, but $\db s_+ =0$ on $S$. Then the extension class is defined by
$\theta_+ = \db s_+/F$.  Similarly, (\ref{eq42}) is represented by
$\theta_- = \db s_-/F$, where $s_-$ is a section of $L^{-2p-k}$ with
the same properties as $s_+$.  It follows that the pull-back of $E$
may be identified with the smooth bundle $C^\infty(p,-p-k)\oplus
C^\infty(-p,p+k)$ with the twisted $\db$-operator
$$ \db_+(u,v) = (\db u+ \theta_+ v,\db v) $$
and equally with $C^\infty(-p-k,p)\oplus
C^\infty(p+k,-p)$ with the twisted $\db$-operator
$$ \db_-(u,v) = (\db u+ \theta_- v,\db v).$$
Now $\theta_+$ is of bidegree $(2p,-2p-2k)$ while $\theta_-$ is of
bidegree $(-2p-2k,2p)$ so the former extends through $L^+$ and the
latter through $L^-$.  It follows that $\pi^*E$ extends to $\IP_3$, as
required.

With this understood, we can try to prove our claim.  $V_r$ maps into
$H^1(Q, \CO(r-2,-k-r)$ which vanishes if $r=1$ and extends through
$L^+$ if $r\geq 2$. On the other hand,
$$
H^0(S,L^r(k-2,0)) =  H^0(S,\CO(r-2p-2,2p+k-r))
\longrightarrow H^1(Q,\CO(r-2p-2-k,2p-r))
$$
injectively for $r=1,2,\ldots, 2p+1$ and again this vanishes identically
if $r=2p+1$ and extends through $L^-$ if $r \leq 2p$.  Hence each
$v\in V_r$ gives rise to an element of $H^1(EL^{r-p}(-2,0))$ which
extends to $\IP_3$ on being pulled back. The claim now follows from
the vanishing theorem of \cite{ADHM}.

\subsection{Boundary conditions}
\label{sec:boundary}
To finish the proof of Theorem \ref{th:BA}
we need to use the boundary conditions satisfied
by the $(A, B, D)$ coming from a monopole \cite{BA}.
There are actually two of these
which are interchanged by the real structure but
for our purposes it is enough to know that  \cite{BA}
$B_1 - D_1 A_1
$ is rank $1$.  Letting $X = B_1 - D_1 A_1 $ we have
$$
M_1(\eta, \zeta) = (\zeta + D_1)(A_1\eta + 1) + X\eta.
$$

We can use this to factorise
$$
M_1 \colon \O^k(-1, -1) \to \O^k
$$
into $M_0(\eta, \zeta) = G(\zeta)\circ F(\eta)$ where
$$
F(\eta) \colon \O^k(-1, -1) \to \O^{k+1}(0, -1)
$$
and
$$
G(\zeta) \colon \O^{k+1}(0, -1) \to \O^k
$$
as follows.  We identify the image of $X$ with $\O$
and then define $F(v)$ by $F(v) = (A_0(v) \eta + v, X(v)\eta)$.
 We define $G$ by
$G(v, w) = \zeta v + D_0(v) + w$ where we identify $\O$ with
the image of $X$ which is inside $\O^k$.

It follows from \cite{At} that the spectral
curve of a monopole does not intersect the anti-diagonal.
In particular it cannot contain a generator of the
quadric.
We claim that this implies that $F$ is injective
and $G$ is surjective. To see this note  that
if $F(\eta)$ is not injective for some $\eta$ then $M_0(\eta, \zeta) =
G(\zeta)F(\eta)$ is not injective for that $\eta$ and all
$\zeta$  so $\det(M_0)$ would
vanish on a generator which  is not possible.
Similarly  $G(\zeta)$ must be onto for all $\zeta$.

Let $J$ be the kernel of $G$ and $V$ the image of $F$. Then
from the discussion in the previous paragraph the only
way that $\det(M_0)$ can vanish is when $J \subset V$.
Hence the
spectral curve is given precisly by the condition $J \subset V$. Moreover
the cokernel of $M_0$ is the cokernel of $G$ and hence we
have
$$
0 \to J \to V \to \O^k \to \coker(M_0) \to 0
$$
as an exact sequence of bundles over the spectral curve. So
we have
$$
\coker(M_0) = \det(V)^* \otimes J.
$$
But we also have that $V = \O^k(-1, -1)$, as $F$ is injective, so
$\det(V) = \O(-k , -k)$. Moreover we have
$$
0 \to K \to \O^{k+1}(0, -1) \to \O^k
$$
and hence
$$
K = \det(\O^{k+1}(0, -1) ) = \O(0, -k-1).
$$
Finally
$$
\coker(M_0) = \O(k, -1).
$$
Now applying Theorem \ref{th3} proves Theorem \ref{th:BA}.

\section{Concluding Remarks}

We have given a rather complete account of the `discrete
linearization' of the discrete Nahm equations on the Jacobian of
algebraic curves in $\IP_1\times \IP_1$.  We have also shown that a
solution of these equations, corresponding by Braam--Austin to a
hyperbolic monopole, arises by a canonical application of our
construction, the algebraic curve in this case being the  spectral
curve of the monopole.

Apart from Conjecture~\ref{conj1}, various questions remain.  In one
direction, it would be of interest to compare our results with other
approaches to discretizations of integrable systems such as
\cite{MV}. One could also ask for an elaboration of the method to
yield explicit (e.g.\ in terms of $\vartheta$-functions) solutions of
the DN system.  This would presumably entail an appropriate analogue
of the methods developed in \cite{ES}.  In this connection we note
that in \cite{W} the general solution of $k=2$ is written down in
terms of elliptic functions (though the boundary conditions are not
considered in detail there).  A special case, is a solution in
trigonometric functions, corresponding to the axially symmetric
hyperbolic monopole with $k=2$, is given by the following.

\begin{example} Pick $p>0$, let $\phi = \pi/(2p+2)$
and
$$
S_p = \{(\eta - e^{i\phi}\zeta)(\eta - e^{-i\phi}\zeta)=0\}.
$$
Then $S_p$ is a real reducible curve in $Q$ and the restriction of
$L^{2p+2}$ to $S$ is holomorphically trivial.  Applying our
construction, with $\CL_1 = \CO(2,-1)$ as in Theorem~\ref{th:BA}, we
obtain the solution
$$
A_r = \begin{pmatrix} 0 & - s/{s_{r+1}}\cr 0 & 0 \cr\end{pmatrix},\;\;
B_r = \begin{pmatrix} -s_{r+1}/s_{r} & 0 \cr 0 & -s_{r}/s_{r+1}
\cr\end{pmatrix},\;\;
D_j = \begin{pmatrix} 0 & 0\cr
s/{s_{r}} & 0 \cr\end{pmatrix},
$$
with
$$
P^+_r = \begin{pmatrix} 1 & 0 \cr 0 & s_{r}/s_{r+1}
\cr\end{pmatrix},\;\;
P^-_r =  \begin{pmatrix} -s_{r+1}/s_r & 0 \cr 0 & -1
\cr\end{pmatrix},
$$
where $s = \sin \phi$ and $s_k = \sin k\phi$.  The solution satisfies
the Braam-Austin boundary condition at $r=1$, and the corresponding
one at $r=2p$ if this is an integer.
In this case, the DN
equations come down to the trigonometric identity
$$
\sin a\phi\, \sin(a+2)\phi + \sin^2 \phi = \sin^2 (a+1)\phi.
$$
The solution is also real ($A_r = - D_r^*$, $B_r = B_r^*$, $P_r^+ = -
[P_{r+1}^-]^*$) with respect to the
hermitian inner product $g_r$ on $V_r= H^0(S,L^r(1,0))$ given by the matrix
$$
g_r = \begin{pmatrix} s_{r+1} & 0 \cr 0 & s_{r} \cr\end{pmatrix}.
$$
Note that these formulae are {\em
algebraic} in the coefficients of $S_p$ as is to be expected.
\end{example}

\vspace {12pt}
We shall now make some remarks which fit our construction in a more
general framework.  First of all, in \S\S2--3 we have closely followed
\cite{NJH2}  and have used `elementary' arguments throughout.
However, the representation of $\CL$ over $S$ as a cokernel is a
special case of the Beilinson spectral sequence, which also gives rise
to monad-type descriptions of classes of holomorphic bundles over
$\IP_2$ and $\IP_3$.

To describe this, consider $Q\times Q$ equipped with its two
projections $p_1$ and  $p_2$. Write $\CO(a,b)(c,d)' =
p_1^*\CO(a,b)\otimes p_2^*\CO(c,d)$
and $\CL'= p_2^*\CL$, viewed as a sheaf on $Q\times Q$. Then we have a
projective resolution of the diagonal $\Delta(Q) \subset Q\times Q$
\be{eq2211}
\CS_{-2} \to \CS_{-1} \to \CS_{0} \to \CO_{\Delta(Q)} \to 0
 \end{equation}
 where
 $$
 \CS_{-2} = \CO(-1,-1)(-1,-1)',\;
 \CS_{-1} = \CO(0,-1)(0,-1)'\oplus \CO(-1,0)(-1,0)',\;
 \CS_{0} = \CO,
 $$
and the maps are given by multiplication by $\langle w,w'\rangle$
and $\langle z,z'\rangle$, $w,z,w',z'$ being the obvious homogeneous
coordinates on the two factors.

Then $p_{1*}(\CL'\otimes \CO_\Delta)$ is
isomorphic to $\CL$.  On the other hand, this direct image is also
computed by the Beilinson spectral sequence
$$
E_1^{ij} = R^jp_{1*}(\CS_{i}\otimes\CL').
$$
The assumption that $\CL$ is regular gives that $E_1^{-1,j}=0$ and it
is also easily checked that $E_1^{-2,0}= E_1^{0,1}=0$. Hence we obtain
$E_1= E_2$ and an exact sequence
$$
0\to H^1(\CL(-1,-1)\otimes\CO(-1,-1) \stackrel{d_2}{\longrightarrow}
H^0(\CL)\otimes \CO
\longrightarrow \CL  \to 0.
$$
Now from (\ref{eq213}) and the assumption that $\CL$ is regular, the
connecting homomorphism $\delta:H^0(S,\CK\otimes\CL) \to
H^1(\CL(-1,-1))$ is an isomorphism.  Finally it can be checked that
that the composite $d_2\circ \delta$ agrees with $\alpha(\CL)$.

\vspace{12pt}
Our last remarks concern the operators $W^{\pm}$ of \S\ref{LAX} that
gave a `Lax representation' for the DN equations; we claim that  these are
essentially the linear operators of the monad description of the
bundle $\tE$ over $\IP_3$.  Indeed a straightforward but tedious
comparison with \cite[\S3]{BA}, shows that their monad (3.1) can be
naturally interpreted as the sequence
$$
\bigoplus_j V_{2j}\otimes L^{-j}(-\frac{1}{2},-\frac{1}{2})
\xrightarrow{(W^+,W^-)}
\bigoplus_j  V_{2j}\otimes(L^{\frac{1}{2}-j}\oplus
L^{-\frac{1}{2}-j})  \xrightarrow{(-W^-,W^+)} \bigoplus_j V_{2j}\otimes
L^{-j}(\frac{1}{2},\frac{1}{2})
$$
down on $Q$. Here we have labelled the vector spaces as in Braam and Austin
because it is more symmetrical, and $W^{\pm}$ are as in
(\ref{eq309}).  The main point we want to make is that the basic
condition that the monad maps form a {\em complex} now becomes the
integrability condition $[W^+,W^-]=0$.  It seems very likely that with
a little further work one should be able to obtain a canonical
identification of $V_{2j}$ here with the space $H^0(S,L^{j+ p
+\frac{1}{2}}(k-1,0))$, thereby giving another proof of
Theorem~\ref{th:BA}, but we shall not pursue this here.  We remark
also that in \cite{npb} the Beilinson spectral sequence is applied to
give monad descriptions of stable bundles over Hirzebruch surfaces and
in particular over $Q$. Those monads are {\em different} from the one
above, and it would be interesting to clarify the relation between
them.

\end{document}